\documentclass[bm]{epl_ARC}

\usepackage{epsfig}
\usepackage{bm}

\begin{document}

\newcommand{\hide}[1]{}
\newcommand{\tbox}[1]{\mbox{\tiny #1}}
\newcommand{\half}{\mbox{\small $\frac{1}{2}$}}
\newcommand{\sinc}{\mbox{sinc}}
\newcommand{\const}{\mbox{const}}
\newcommand{\trc}{\mbox{trace}}
\newcommand{\intt}{\int\!\!\!\!\int }
\newcommand{\ointt}{\int\!\!\!\!\int\!\!\!\!\!\circ\ }
\newcommand{\eexp}{\mbox{e}^}
\newcommand{\bra}{\left\langle}
\newcommand{\ket}{\right\rangle}
\newcommand{\EPS} {\mbox{\LARGE $\epsilon$}}
\newcommand{\ttimes} {\mbox{\tiny \ $^{\times}$ \ }}
\newcommand{\ar}{\mathsf r}
\newcommand{\im}{\mbox{Im}}
\newcommand{\re}{\mbox{Re}}
\newcommand{\bmsf}[1]{\bm{\mathsf{#1}}} 
\newcommand{\ssum}{\sum}
\newcommand{\mpg}[3][b]{\begin{minipage}[#1]{#2}{#3}\end{minipage}}


\title{The information entropy of quantum mechanical states}

\shortauthor{QM information entropy}
\shorttitle{}

\author{\small
Alexander Stotland\inst{1}, 
Andrei A. Pomeransky\inst{2}, 
Eitan Bachmat\inst{3}
and Doron Cohen\inst{1}
}

\institute{
\inst{1} {\small Department of Physics, Ben-Gurion University, Beer-Sheva 84105, Israel} \\
\inst{2} {\small Laboratoire de Physique Th\'eorique, UMR 5152 du CNRS, Universit\'e Paul Sabatier, 31062 Toulouse Cedex 4, France} \\
\inst{3} {\small Department of Computer Science, Ben-Gurion University, Beer-Sheva 84105, Israel}
}


\maketitle


\begin{abstract}
It is well known that a Shannon based definition of information 
entropy leads in the classical case to the Boltzmann entropy.  
It is tempting to regard the Von Neumann entropy as the corresponding 
quantum mechanical definition. But the latter is problematic 
from quantum information point of view. Consequently we introduce 
a new definition of entropy that reflects the inherent uncertainty 
of quantum mechanical states. We derive for it an explicit expression, 
and discuss some of its general properties. We distinguish between 
the minimum uncertainty entropy of pure states, 
and the excess statistical entropy of mixtures. 
\end{abstract}



The statistical state of a system ($\rho$) is specified in classical 
mechanics using a probability function, while in the quantum 
mechanical case it is specified by a probability matrix. 
The information entropy $S[\rho]$ is a measure for the amount 
of extra information which is required in order to predict 
the outcome of a measurement. If no extra information is needed 
we say that the system is in a definite statistical state with $S=0$. 
A classical system can be in principle prepared in a definite 
state. But this is not true for a quantum mechanical system. 
Even if the system is prepared in a pure state, still there is 
an inherent {\em uncertainty} regarding the outcome of a general measurement. 
Therefore the minimum information entropy of a quantum mechanical state  
is larger than zero.

It is clear that the common von Neumann definition of quantum mechanical 
entropy does not reflect the inherent uncertainty which is associated 
with quantum mechanical states \cite{motive1,motive2}. 
For a pure state it gives $S=0$. 
Let us assume that we prepare two spins in a (pure) singlet state. 
In such a case the von Neumann entropy of a single spin is $S=\ln(2)$,  
while the system as a whole has $S=0$. If it were meaningful to give 
these results an information theoretic interpretation, it would be implied 
that the amount of information which is needed to determine the 
outcome of a measurement of a subsystem is larger than the amount 
of information which is required in order to determine the outcome 
of a measurement of the whole system. This does not make sense.

Thus we are faced with the need to give a proper definition 
for the (information) entropy of a quantum mechanical state.  
As in the case of the von Neumann entropy it can be regarded 
as a measure for the lack of purity of a general (mixed) state. 
But unlike the von Neumann entropy it does not give $S=0$ 
for pure states, and does not coincide with the thermodynamic 
entropy in case of a thermal state.

In this Letter we introduce a Shannon-based definition 
of quantum mechanical information entropy; derive explicit 
expressions for the calculation of this entropy; 
and discuss some of its properties. For further motivations 
and review of the traditional definition of entropy 
in the context of quantum computation and quantum information 
see \cite{book}.

The statistical state of a classical system,  
that can be found in one of $N$ possible 
states $r$, is characterized by the corresponding 
probabilities $p_r$, with the normalization $\sum p_r =1$. 
The amount of information which is required in order 
to know what is going to be the outcome of a measurement 
is given by the Shannon formula:
\mbox{${\cal S} = -\sum_r p_r \ln(p_r)$}.
Note that $S=0$ if the system is in a definite state,
while $S=\ln(N)$ in the worst case of a uniform distribution. 
This definition coincides with the Boltzmann definition of 
entropy if $r$ are regarded as phase space cells.

In the quantum mechanical case the statistical state
of a system is described by a probability matrix $\rho$.
A measurement requires the specification of a basis
of (pure) states $|a\rangle$. Without any loss of generality
it is convenient to define a given basis by specifying
a hermitian operator ${\cal A}$. 
We note that in a semiclassical context the basis ${\cal A}$ 
can be regarded as a {\em partitioning} of phase space 
into cells.  
The probability to have $a$ as the outcome of a measurement
is $\langle a | \rho | a \rangle$.
Therefore the information entropy for such a measurement is
\begin{eqnarray} \label{e2}
S[\rho|{\cal A}] = -\sum_a \langle a | \rho | a \rangle   \ln(\langle a | \rho | a \rangle)
\end{eqnarray}
Our notation emphasizes that this is in fact a
conditional entropy: one has to specify in advance 
what is the measurement setup.  
In particular there is a basis ${\cal H}$ in which $\rho$ 
is diagonal $\rho=\mbox{diag}\{p_r\}$. In this basis 
$S[\rho|{\cal A}]$ attains its {\em minimum} value   
\begin{eqnarray} \label{e_4}
S_{\tbox{H}}[\rho] = S[\rho|{\cal H}] =  -\sum_r p_r \ln(p_r) = -\trc(\rho\ln\rho)
\end{eqnarray}
which is known as the von Neumann entropy.
We would like to emphasize that from {\em strict} information 
theory point of view, the quantity $S_{\tbox{H}}[\rho]$ 
can be interpreted as information entropy (a la Shanon) 
only if we assume a-priori knowledge of 
the preferred basis that makes $\rho$ diagonal.
In equilibrium statistical mechanics the interest is 
in stationary states. This means that $\rho$
is diagonal in the basis that is determined by the
Hamiltonian ${\cal H}$. Therefore, if we measure the energy
of the system, the information entropy 
is indeed $S_{\tbox{H}}[\rho]$. 
In particular for a canonical state $\rho \propto \exp(-\beta {\cal H})$
it reduces to the thermodynamic definition of entropy.

For a pure quantum mechanical state $\rho=|\Psi\rangle\langle\Psi|$  
the von Neumann definition gives \mbox{$S_{\tbox{H}}[\rho]=0$}. 
This seems to imply that a pure quantum mechanical
state is lacking a statistical nature. This is of course not correct.
For a general  measurement we have {\em uncertainty}.
An absolute definition of an information entropy
of a quantum mechanical state should not assume
any special basis. This imply a {\em unique} definition 
of the {\em absolute} entropy. Using standard information theory   
argumentation we conclude that 
\footnote{
If we regard the measurement apparatus as a part of the 
system, then information theory tells us that the total entropy is 
$S_{\tbox{total}}=S[{\cal A}]+\sum_{\cal A} P({\cal A}) S[\rho|{\cal A}]$. 
The probability $P({\cal A})$ describes our lack of 
knowledge regarding the state of the apparatus, 
and $S[A]$ is its corresponding entropy. 
Quantum mechanics assumes that there is no preferred basis.}
\begin{eqnarray} \label{e3}
S[\rho] 
\ = \       \overline{S[\rho|{\cal A}]}
\ = \       S_0(N) + F(p_1,p_2,...) 
\ \equiv \  S_0(N) + S_{\tbox{F}}[\rho]
\end{eqnarray}
where the overline indicates averaging over all possible
basis sets with uniform measure (no preferred basis).
We would like to emphasize that the averaging procedure 
is {\em unique}: A choice of a basis is like a choice
of ``direction" in a $2N-1$ dimensional space 
(in the case of spin $1/2$ this direction can 
be interpreted as the geometrical orientation 
of our $xyz$ axes in the physical space).       
The second equality in Eq.(\ref{e3}), gives an explicit 
expression for the absolute entropy, which we are going 
to derive below.  The result is written as a sum of two terms:  
The first term is the {\em minimum uncertainty entropy} of 
a quantum mechanical state, achieved by a pure state, 
while the second term gives 
the deviation from purity.  We shall call the second term 
excess statistical entropy, and will use for it 
the notation $S_{\tbox{F}}[\rho]$.  Conceptually it 
is meaningful to ask to what extent 
$S_{\tbox{F}}[\rho]$ is correlated with $S_{\tbox{H}}[\rho]$. 
We shall discuss this issue later on.

Assume that $\rho=\mbox{diag}\{p_r\}$ is diagonal 
in some basis ${\cal H}$. We can regard all 
the possible ${\cal A}$ basis sets, as unitary ``rotations" 
of ${\cal H}$. This means that any $a \in {\cal A}$ 
in the rotated basis is obtained from a state $r \in {\cal H}$ 
in the preferred basis by an operation $U$. 
Consequently 
\begin{eqnarray} \label{e4}
S &=& \overline{\sum_{a} f \left(\sum_{r} p_{r} |\langle r | a \rangle|^2\right)}^{\ A}  
= \overline{\sum_{s} f \left(\sum_{r} p_{r} |\langle r |U| s \rangle|^2\right)}^{\ U} 
\\ \label{e6}
&=& N \overline{f\left(\sum_{r} p_{r}|\langle r| \Psi \rangle |^2\right)}^{\ \Psi} 
= N \overline{f\left(\sum_{r} p_{r}(x_r^2 + y_r^2)\right)}^{\ \tbox{sphere}} 
= N \int_0^{\infty} f(s) \ P(s)ds 
\end{eqnarray}
where we use the notation $f(s)=-s\ln(s)$.  
Each averaged $|\langle r |U| s \rangle|^2$ in Eq.(\ref{e4}) 
is equal to $|\langle r| \Psi \rangle |^2$ averaged  
over all possible $\Psi$, which leads to Eq.(\ref{e6}). 
It is important to re-emphasize that the 
quantum mechanical ``democracy" uniquely defines 
the measure for this $\Psi$ average. This becomes 
more transparent if we define $x_r$ and $y_r$ 
as the real and imaginary parts of  $\Psi_r = \langle r | \Psi \rangle$.  
The normalization condition is $\sum_r (x_r^2+y_r^2)=1$.  
Hence in the final expression the average is over 
all possible directions in a $2N-1$ dimensional space. 
In the final expression we introduce the notation 
\begin{eqnarray}
s \ = \ \sum_{r} p_{r} |\Psi_r|^2 = \sum_r p_r (x_r^2 + y_r^2) 
\end{eqnarray}
and its probability distribution is denoted $P(s)$.
In what follows we discuss the calculation of $P(s)$ 
and its integral with $f(s)$.

In case of a maximally mixed state $P(s)$ is 
delta distributed around $s=1/N$, 
and hence $f(s) = \ln(N)/N$.  The 
corresponding information entropy 
is therefore $S[\rho]=\ln(N)$ as expected. 
If the state is not maximally 
mixed then $P(s)$ becomes non trivial.  
In case of a pure state $s=|\Psi_1|^2$ 
and its distribution is well know \cite{haake}:  
\begin{eqnarray} \label{e9}
P(s) = (N-1)(1-s)^{N-2}
\end{eqnarray}
Thus we get an expression for the 
{\em ``minimum uncertainty entropy"} 
which is $N$ dependent: 
\begin{eqnarray} \label{e10}
S_0(N) \ = \ \sum_{k=2}^{N} \frac{1}{k} 
\ \approx \ \ln(N) - (1{-}\gamma) + \frac{1}{2N}
\end{eqnarray}
Using the asymptotic approximation in the last equality 
we see that the difference between the $S$ of a maximally 
mixed state, and that of a pure state, approaches a universal 
value $(1-\gamma)$, where $\gamma$ is Euler's constant.
Using different phrasing, we see that the excess 
statistical entropy is universally bounded:
\begin{eqnarray} 
S_{\tbox{F}}[\rho] \ < \ 1-\gamma
\end{eqnarray}

To get an actual expression for the excess statistical entropy, 
due to lack of purity, requires some more effort. 
The first stage is to calculate $P(s)$ leading to (see appendix):
\begin{eqnarray} \label{e11}
P(s)= (N{-}1) \sum_{(p_r > s)} 
\left[\prod_{{r'}(\neq r)} \frac{1}{p_{r}{-}p_{r'}}\right] (p_r-s)^{N-2}  
\end{eqnarray}
The second stage is to calculate the integral 
of Eq.(\ref{e6}) using
\begin{eqnarray} \nonumber
\int_0^p (p{-}s)^{N-2} s \ln(s)ds = \frac{p^N}{N(N-1)}
\left[\ln(p) - \sum_{k=2}^n \frac{1}{k}\right]
\end{eqnarray}
and then to use the identity (see appendix)
\begin{eqnarray} \label{eid1}
\sum_r p_r^N \prod_{{r'}(\neq r)} \frac{1}{p_{r}{-}p_{r'}} 
\ \ = \ \ \sum_r p_r \ \ = \ \ 1
\end{eqnarray}
Hence one obtains
\begin{eqnarray} \label{e12}
F(p_1,p_2,...) = - \sum_r \left[\prod_{r' ( \ne r)} \frac{p_r}{p_r-p_{r'}} \right] 
\ p_r \ln(p_r) 
\end{eqnarray}
This expression is independent of $N$. 
Namely, extra zero eigenvalues 
do not have any effect on the result.  
Some particular cases are of interest.
For a mixture of two states we get
\begin{eqnarray} \label{e13}
F(p_1,p_2) = -\frac{1}{p_1-p_2}(p_1^2\ln(p_1)-p_2^2\ln(p_2))
\end{eqnarray}
For a uniform mixture of $n$ states we get 
\begin{eqnarray} \label{e14}
S_{\tbox{F}}[\rho] &=& \ln(n) - \sum_{k=2}^n \frac{1}{k} 
\\ \label{e14b}
S[\rho] &=& \ln(n) + \sum_{n < k \le N} \frac{1}{k}
\end{eqnarray}

Either $S_{\tbox{H}}[\rho]$ or $S_{\tbox{F}}[\rho]$ 
can serve as a measure for lack of purity. 
In Fig.1 we present results of calculation of $S_{\tbox{F}}[\rho]$ 
versus $S_{\tbox{H}}[\rho]$ for a set of representative 
states, both uniform and non-uniform mixtures. 
We see that there is a very strong correlation between 
these two (different) measures of purity.

Our definition of entropy has some interesting 
mathematical properties. One simple property is concavity: 
Given $0<\lambda<1$ and two sets of probabilities we have  
\begin{eqnarray} 
F(\lambda p_r + (1-\lambda)q_r) \ge \lambda F(p_r) + (1-\lambda) F(q_r)
\end{eqnarray}
This follows from the concavity of $f(s)$ in Eq.(\ref{e6}). 
Concavity and symmetry with respect to the variables $p_i$ imply
that $S[\rho]$ attains its maximum for maximally mixed states 
and its minimum for pure states. 
This property is helpful for justifying argumentations  
that are based on ``worst case" calculations. 
Below we list some less trivial properties 
which are of physical interest.

Consider a system in a state $\rho$, 
and its subsystem which is in some state $\sigma$. 
Technically the reduced probability matrix $\sigma$ 
is obtained from $\rho$ by tracing over the irrelevant
indexes. From general information theoretic 
considerations we expect 
\begin{eqnarray}  \label{ei1}
S[\sigma] < S[\rho]
\end{eqnarray}
This means that determination of a state 
of a subsystem requires less information.
As explained in the introduction this inequality 
is violated by the von Neumann entropy. 
But with our definition 
$S[\rho] \ge S_0(MN) \ge S_0(2N) > \ln(N) > S[\sigma]$, 
where $N$ and $MN$ are the dimensions 
of $\sigma$ and $\rho$ respectively.

Another common physical situation is  
having a state $\rho = \sigma_{\tbox{A}} \otimes \sigma_{\tbox{B}}$ 
where $\sigma_{\tbox{A}}$ 
and $\sigma_{\tbox{B}}$ are states of subsystems that 
were prepared independently(!). 
Obviously we have the property  
\begin{eqnarray} 
S[\rho|{\cal A}\otimes{\cal B}]
=S[\sigma_{\tbox{A}}|{\cal A}]+S[\sigma_{\tbox{B}}|{\cal B}]
\end{eqnarray}
But for the absolute information entropy we expect 
\begin{eqnarray}  \label{ei2}
S[\rho] \ge S[\sigma_{\tbox{A}}] + S[\sigma_{\tbox{B}}] 
\end{eqnarray}
This comes about because there are bases 
which are not ``external tensor product"  
of ${\cal A}$-basis and ${\cal B}$-basis.
Thus this inequality reflects the greater 
uncertainty that we have in the state determination 
of the combined system. Note that if our world were 
classical, we would get an equality, which is 
the case with the Boltzmann entropy, and in fact also 
with the von Neumann entropy.
In order to better establish Eq.(\ref{ei2})
we can consider a worst case scenario. 
Let $N$ and $M$ be the dimensions 
of $\sigma_{\tbox{A}}$ and $\sigma_{\tbox{B}}$ respectively. 
Assume that these states are uniform 
mixtures of $n$ and $m$ states respectively, then
$\rho$ is a uniform mixture of $nm$
states in dimension $NM$. 
Using Eq.(\ref{e14b}) and the inequality 
{\small
\begin{eqnarray} \nonumber
\sum_{k=nm+1}^{NM} \frac{1}{k}  
= \sum_{k=nm+1}^{mN} \frac{1}{k} 
+ \!\!\!\sum_{k=mN+1}^{NM} \frac{1}{k} 
= \sum_{k_1=n+1}^N \sum_{l_1=0}^{m-1}\frac{1}{k_1 m{-}l_1}
+ \!\!\!\sum_{k_2=m+1}^M\sum_{l_2=0}^{N-1} \frac{1}{k_2 N{-}l_2} 
> \sum_{k=n+1}^N \frac{1}{k}
+ \!\!\!\sum_{k=m+1}^{M}\frac{1}{k} 
\end{eqnarray}
}
we confirm that Eq.(\ref{ei2}) is indeed satisfied.
%
%
%
%
%
%
A particular case of the inequality of Eq.(\ref{ei2}) 
is that the minimum uncertainty entropy satisfies 
\begin{eqnarray} \label{ei3a}
S_0(NM) > S_0(N) + S_0(M)
\end{eqnarray}
But what about the excess statistical entropy? 
Our conjecture is that 
\begin{eqnarray} \label{ei3}
S_{\tbox{F}}[\rho] \le S_{\tbox{F}}[\sigma_{\tbox{A}}] + S_{\tbox{F}}[\sigma_{\tbox{B}}] 
\end{eqnarray}
We can again establish this inequality 
for uniform mixtures of $n$ and $m$ states 
in dimensions $N$ and $M$ respectively: 
Using Eq.(\ref{e14}) we observe that 
\begin{eqnarray} \nonumber
S_{\tbox{F}}[\rho]-S_{\tbox{F}}[\sigma_{\tbox{A}}]-S_{\tbox{F}}[\sigma_{\tbox{B}}] = 
S_0(n)+S_0(m)-S_0(nm)
\end{eqnarray}
which is negative by Eq.(\ref{ei3a}). 
It is important to realize that Eq.(\ref{ei3a}) over compensates 
the inequality Eq.(\ref{ei3}) leading to Eq.(\ref{ei2}). 
It is well known that for the von Neumann entropy 
we have the general inequality
\begin{eqnarray} 
S_{\tbox{H}}[\rho] \le S_{\tbox{H}}[\sigma_{\tbox{A}}] + S_{\tbox{H}}[\sigma_{\tbox{B}}]
\end{eqnarray}
which holds for any subdivision of a system into 
two (correlated) subsystems.  We already observed (Fig.~1) 
that $S_{\tbox{F}}[\rho]$ is strongly correlated 
with $S_{\tbox{H}}[\rho]$. Moreover, this correlation 
is {\em sublinear}. It follows that we expect 
the easier inequality Eq.(\ref{ei3}) to hold in general, 
also in case of correlated subsystems.

The effect of quantum measurements 
on the entropy is of special interest. 
Let $P_i$ be a complete orthogonal set 
of projectors ($\sum_iP_i=1$). 
The state after a projective measurement 
is $\sigma = \sum_i P_i \rho P_i$.  
Consequently the state of the system 
becomes more mixed. This is indeed  
reflected by an increase in the  
von Neumann entropy of the systems. 
Also our entropy is a measure for lack of purity.  
Therefore it is reasonable to expect 
$S[\sigma] \geq S[\rho]$. We were not 
able to prove this assertion.

\vspace*{3mm}

{\bf Summary:} 
The von Neumann entropy $S[\rho|{\cal H}]$ is useful 
in the thermodynamic context, where the interest is 
a-priori limited to stationary (equilibrium) states. 
If we want to study the growth of entropy during 
an ergodization process, we may consider $S[\rho|{\cal A}]$, 
where ${\cal A}$ is a basis (or a ``partition" of phase space) 
that does not commute with ${\cal H}$. 
See for example Ref.\cite{piotr} where entropy 
is defined with respect to the position representation. 
In the latter case the entropy of a pure state is 
in general non-zero. 
In the present study we have derived an explicit expression  
for the minimum uncertainty entropy $S_0(N)$ of pure 
states. This can be associated with the {\em average}   
over the minimum entropic uncertainty \cite{uncert}.      
We also have derived an expression for the excess statistical 
entropy $S_{\tbox{F}}[\rho]$ of mixtures. 
The latter can be used as a measure for lack 
of purity of quantum mechanical states, and it is strongly 
correlated with the von Neumann entropy $S_{\tbox{H}}[\rho]$. 
It is bounded from above by $(1-\gamma)$, where $\gamma$ 
is Euler's constant. The total information entropy $S[\rho]$, 
unlike the von Neumann entropy, has properties that 
do make sense from quantum information point of view.

\vspace*{3mm}

{\bf Appendix:}
Switching to the variables $s_r=x_r^2+y_r^2$ 
the definition of $P(s)$ takes the form
\begin{eqnarray} \nonumber
P(s) = \left\langle \delta(s-\sum_r p_r (x_r^2+y_r^2))\right\rangle_{\tbox{sphere}} 
= \ (N{-}1)! \int_0^\infty \!\! ds_1..ds_N  
\ \delta(1{-}\ssum_r s_r) \delta(s{-}\ssum_r p_r s_r)
\\ \nonumber 
= (N{-}1)! \int_0^{\infty} \!\!\! ds_1..ds_N \int\frac{d\omega d\nu}{(2\pi)^2}
\eexp{(1{-}\ssum_r s_r)(i\nu+0) + i(s{-}\ssum_r p_r s_r)\omega} 
\end{eqnarray}
where the infinitesimal $0$ has been introduced to insure 
convergence once the order of integration is changed.
Thus after the integration over $ds_1...ds_N$ one has
\begin{eqnarray} \nonumber
P(s) = (N{-}1)! \int \frac{d\omega d\nu}{(2\pi)^2} 
\eexp{i\nu+i\omega s}
\prod_r \frac{1}{i\omega p_r + i\nu + 0}
= \int \frac{d\omega}{2\pi} \frac{(N{-}1)!}{(i\omega)^{N{-}1}}
\ssum_r \eexp{i\omega (s-p_r)}\prod_{r' (\neq r)} \frac{1}{p_{r'}-p_{r}}
\end{eqnarray}
One can show (see below) that there is no singularity in the integral 
at $\omega=0$, so one can deform the contour 
of integration in such way that it will go slightly 
above the point $\omega=0$. Then one can make the integral 
term by term leading to the final result Eq.(\ref{e11}).
Namely, if $p_r<s$ the contour is closed in the upper 
half plan leading to zero, while if $p_r>s$ 
the contour is closed in the lower half plan 
leading to a non-zero contribution from the $\omega=0$ pole.

The above manipulation was based on the observation that 
the integrand {\em as a whole} is non-singular:  
The $1/\omega^{N{-}1}$ singularity of the individual terms 
cancel upon summation over $r$. This cancellation can be established 
by expanding the exponent in powers of $\omega$, 
and using the identity
\begin{eqnarray} \nonumber
\sum_r (s-p_r)^n \prod_{r^\prime(\neq r)}\frac{1}{p_{r}-p_{r'}} =0
\ \ \ \ \ \mbox{for $n\leq (N{-}2)$} 
\end{eqnarray}
Both this identity and also Eq.(\ref{eid1}) 
can be proved by the following procedure:
\begin{eqnarray}\nonumber
\sum_r g(p_r) \prod_{k (\ne r)} \frac{1}{p_r-p_k} \ = \
\oint \frac{dz}{2\pi i} g(z) \prod_k \frac{1}{z-p_k} 
\ = \ \oint \frac{dz}{2\pi i} z^{N{-}2}g(1/z) \prod_k \frac{1}{1-p_k z}
\end{eqnarray}
where in the last step one changes $z\mapsto 1/z$.

\vspace*{3mm}

DC thanks Thomas Dittrich (Bogota) and Alex Gersten (Beer Sheva) 
for a conversation that had motivated this work, 
and Dima Shepelyansky for an exciting visit in Toulouse. 
We also thank Piotr Garbaczewski (Zielona Gora) for helpful comments. 
This research was supported by the Israel Science Foundation (grant No.11/02).
The work of AP was supported by the NSA and ARDA under ARO
contract No. DAAD19-01-1-0553.




\noindent
\begin{picture}(250,250)
\put(0,0){\epsfig{figure=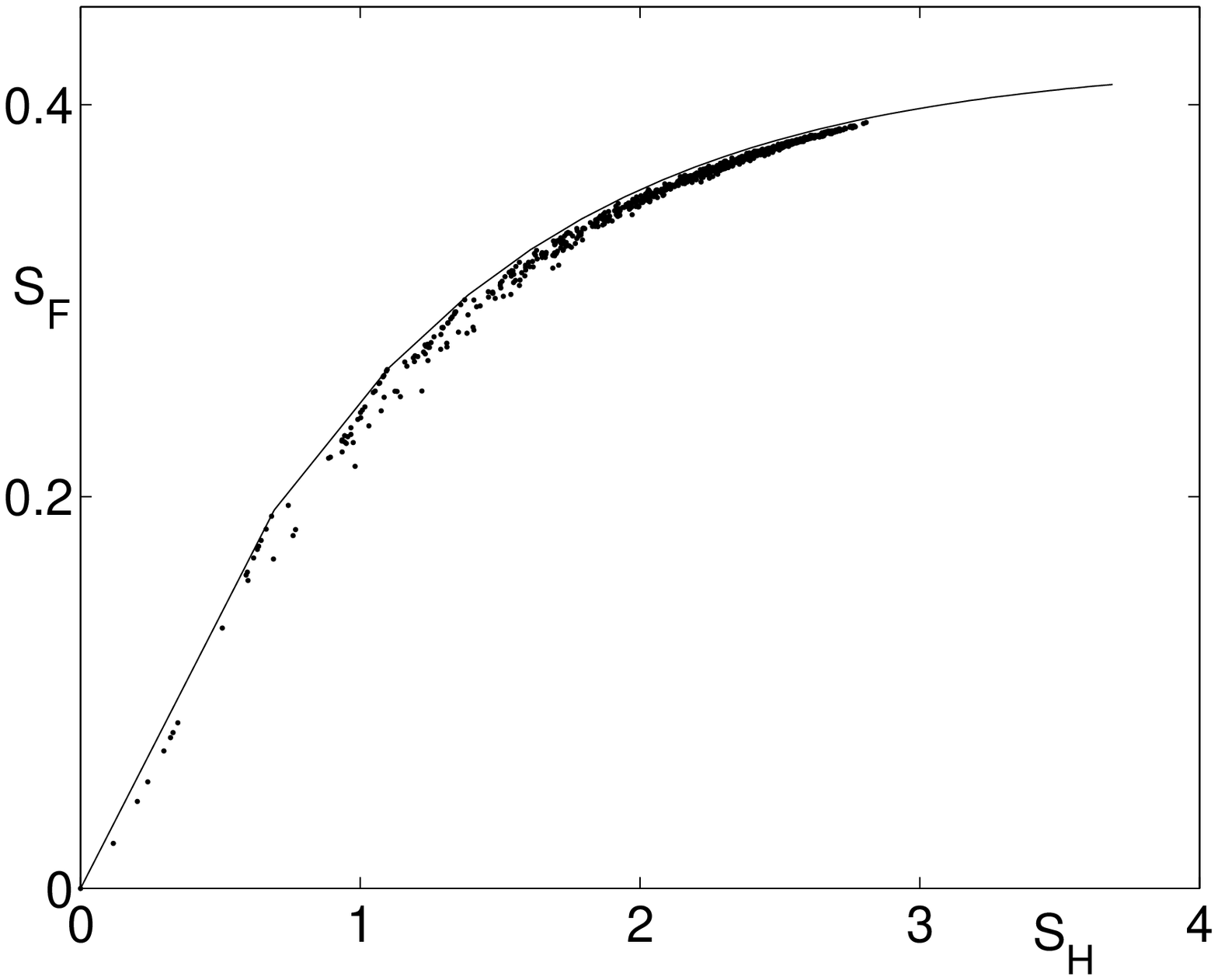,width=90mm}}
\put(120,30){\epsfig{figure=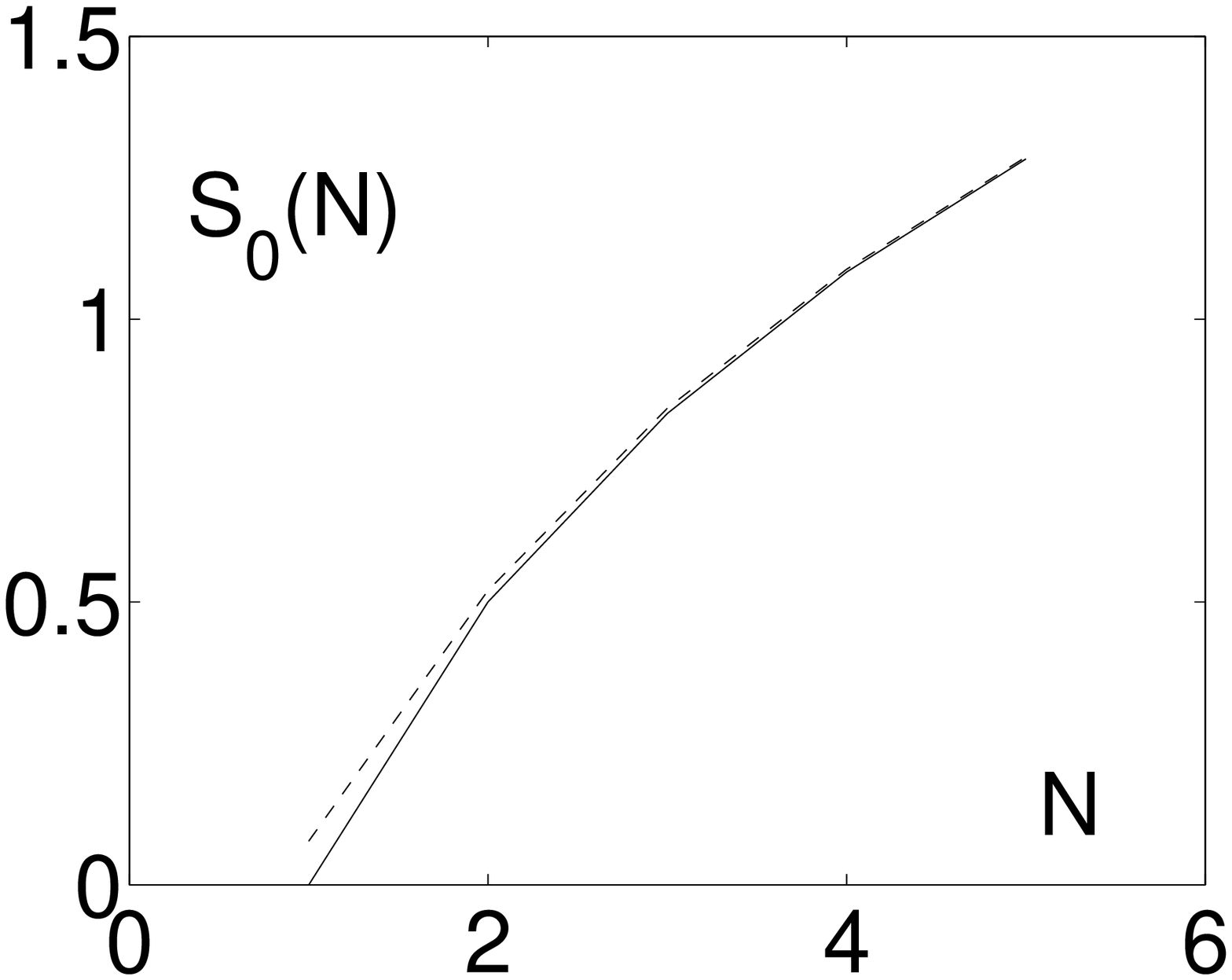,width=45mm}}
\end{picture}

{\footnotesize {\bf Fig.1:} The excess information entropy of a mixed quantum 
mechanical state $S_{\tbox{F}}$ versus the von Neumann entropy $S_{\tbox{H}}$.  
The solid line is for uniform mixtures, while the dots are for randomly 
chosen (nonuniform) mixtures. {\bf Inset:} The information entropy of 
a pure quantum mechanical state as a function of the Hilbert space dimension $N$. 
See Eq.(\ref{e10}). The dashed line is the asymptotic approximation.}


\end{document}